\documentclass{rspublic}
\usepackage{graphicx}
\usepackage[dvips]{epsfig}

\def\R{{\bf R}}
\def\Z{{\bf Z}}
\def\S{{\bf S}}

\begin{document}

\title{Highly-symmetric travelling waves in pipe flow}

\author[C.C.T. Pringle, Y. Duguet, R.R. Kerswell]{Chris C.T. Pringle$^\dag$,
Yohann Duguet$^{\dag*}$ and Rich R. Kerswell$^\dag$}

\affiliation{$^{\dag}$Department of Mathematics, University of Bristol, United Kingdom.
\\$^*$ Linn\'{e} Flow Center, KTH Mechanics, SE-10044 Stockholm, Sweden.}

\label{firstpage}
\maketitle

\begin{abstract}{Pipe flow; travelling waves; turbulence transition}

The recent theoretical discovery of finite-amplitude travelling
waves in pipe flow has re-ignited interest in the transitional
phenomena that Osborne Reynolds studied 125 years ago. Despite all
being unstable, these waves are providing fresh insight into the
flow dynamics. Here we describe two new classes of {\em
highly}-symmetric travelling waves (possessing rotational,
shift-\&-reflect and mirror symmetries) and report a new family of
mirror-symmetric waves which is the first found in pipe flow {\em
not} to have shift-\&-reflect symmetry. The highly-symmetric waves
appear at lower Reynolds numbers than the originally-discovered
non-mirror-symmetric waves found by Faisst \& Eckhardt 2003 and
Wedin \& Kerswell 2004 and have much higher wall shear stresses. The
first {\em M-class} comprises of the various
discrete-rotationally-symmetric analogues of the mirror-symmetric
wave found in Pringle \& Kerswell (2007) and have a distinctive
double layer structure of fast and slow streaks across the pipe
radius. The second {\em N-class} has the more familiar separation of
fast streaks to the exterior and slow streaks to the interior and
looks the precursor to the class of non-mirror-symmetric waves
already known.

\end{abstract}

\section{Introduction}

The stability of the flow of a Newtonian fluid such as water through
a straight pipe of constant circular cross-section has fascinated
scientists ever since Reynolds' (1883) original experiments.
Reynolds realised that there was a single non-dimensional control
parameter for the flow, $Re:= UD/\nu$ (where $U$ is the mean
velocity, $D$ is the pipe diameter, and $\nu$ is the kinematic
viscosity of the fluid), but that there was no unique value of this
parameter for transition to occur. Instead, he noticed that the
value of $Re$ required for transition depended critically on how
cleanly he performed the experiment, that is, on the ambient level
of noise. Later theoretical and computational work confirmed that
the steady, unidirectional, parabolic, laminar `Hagen-Poiseuille'
flow (Hagen 1839, Poiseuille 1840) realised at low $Re$  is linearly
stable at least to $Re=10^7$ (Meseguer \& Trefethen 2003) and that
pipe flow transition must be a finite-amplitude process. The
variability in the critical $Re$ across experimental realisations
has indicated that this process is also sensitive to the exact form
of disturbances present in the flow (e.g. Peixinho \& Mullin 2007).

Once triggered, transition is generally abrupt and can lead to both
localised and global forms of turbulence depending on $Re$. For
$1760 \lesssim Re \lesssim 2,300$, turbulence is localised into
`puffs' (Wygnanski \& Champagne 1973) which have a weak front, a
sharp trailing edge, and maintain their length at about $20\,D$ as
they propagate along the pipe. For $2,300 \lesssim Re$, the
turbulence delocalises into `slug' turbulence (Wygnanski {\em et al}
1975) which has sharp front travelling faster than the mean flow and
a trailing edge travelling slower so that it expands while
propagating downstream. As a result of this varied behaviour and the
sensitivity of  the transition process, pipe flow has become {\em
the} canonical example of shear flow transition.

From a dynamical systems viewpoint, the existence of turbulence
requires the presence of simpler solutions of the governing
equations to provide a framework in state space to sustain turbulent
trajectories from simply relaminarising. Natural questions are then
whether the emergence of such alternative solutions can be used to
predict the threshold (lowest) $Re$ for transition and what role
they play in the process as well as the final turbulent state. 
Steady states and travelling waves (TWs) - fixed points in
an appropriately chosen  Galilean frame - are the simplest
solutions.

The first step towards identifying such states was taken by Smith \&
Bodonyi ($1982$) who predicted the existence of helical modes (with
swirl) through a critical layer analysis. Landman ($1990$), however,
failed to find any numerical evidence for their existence. In the
same year, Nagata (1990) used a homotopy approach to find finite
amplitude solutions in plane Couette flow starting from the
linearly-unstable rotating situation. A similar attempt using
rotating pipe flow, however, failed to reach the vanishing rotation
limit (Barnes \& Kerswell $2000$). By this time a more systematic
approach was emerging based upon the idea of a self-sustaining
process. Hamilton \emph{et al.} ($1995$) suggested that streamwise
rolls could create streamwise streaks which in turn could reinforce
the original rolls by some unspecified non-linear process. Waleffe
($1997$,$1998$) closed the loop by showing that the streaks had
inflectional instabilities which led to what he described as
`wriggling' in such a manner that  energy is fed back into the
initial rolls. He then identified such solutions in plane-Couette
flow using a constructive homotopic approach. Faisst \& Eckhardt
($2003$) and Wedin \& Kerswell ($2004$) employed this technique
within the geometry of pipe flow to find the expected travelling
waves although all were without swirl and therefore unrelated to the
predictions of Smith \& Bodonyi ($1982$). Experimental observations
(Hof et al. 2004,2005) and numerical computations (Kerswell \& Tutty 2007, 
Schneider et al. 2007a, Willis \& Kerswell 2008) then followed which indicated that
these waves are realised albeit transiently as coherent structures
in turbulent pipe flow. These waves are all born in saddle node
bifurcations with the lowest at $Re=1251$ for TWs with a discrete
rotational symmetry. Pringle \& Kerswell
($2007$) substantially lowered this to $773$ by finding travelling
waves with no rotational symmetry. The $Re$ gap between when
alternative solutions exist and when sustained turbulence occurs is
then certainly positive but also too large to be a useful predictor.

Recently Duguet et al (2008a) have adopted an entirely different
approach to finding TWs  based upon identifying episodes in
time-dependent pipe flow which exhibit temporal periodicity. To
generate a long time signal for this search, sustained flow dynamics
away from the laminar state is required. The obvious source for this
is studying flow on the turbulent attractor but they realised that
there was a more promising alternative where the flow is less
energetic and periodic episodes are more clearly evident: the
laminar-turbulent boundary. This is the set of flows which separates
initial conditions which immediately relaminarise from those which
undergo a turbulent episode. An initial condition chosen on this
boundary or `edge' (Skufca et al. 2006) will evolve in time forever
treading a tight rope between relaminarising and becoming turbulent
(Toh \& Itano 1999, Itano \& Toh 2001, Schneider et al 2007b). Duguet
et al (2008a) identified 3 new TWs by using velocity fields from
near-time periodic episodes in this energetically intermediate flow
dynamics as starting guesses for an iterative Newton-Krylov solver.

The purpose of this paper is to show how the TWs found by Duguet et
al (2008a) led to the discovery of 2 new {\em classes} of
mirror-symmetric TWs hereafter referred to as the M- and N-classes,
and a distinctly new family of mirror-symmetric TWs. Each class is
partitioned by the discrete rotational symmetry (see (\ref{R})
below) enjoyed by the waves which defines a {\em family} within the
class. This suggests a natural labelling system where, for example,
the family of M-class TWs with $m$ (integer) degree of discrete
rotational symmetry will be referred to as the M$m$ family. We argue
that the mirror-symmetric wave of Pringle \& Kerswell (2007) is
actually part of the first family (M1) of the M-class  and `A3' in
the nomenclature of Duguet et al (2008a) is a member of the second
family (M2). We further report the families M3, M4, M5 and M6 which,
unlike anything seen so far, all have a double layer structure of
fast and slow streaks across the pipe radius. The second N-class of
TWs has `C3' in Duguet et al (2008a) as  part of its second family
(N2) with the families N3, N4 and N5 described here for the first
time. In these waves, the fast streaks are positioned near the pipe
wall  and slow streaks in the interior, a separation familiar from
the original  non-mirror symmetric TWs found by Faisst \& Eckhardt
(2003) and Wedin \& Kerswell (2004). Since these waves only have
 shift-\&-reflect symmetry - see (\ref{S}) below - we refer to this
 original set as the S-class hereafter for convenience.

The plan of the paper is to start by discussing the various
symmetries of the TWs which serve to categorize them in section 2.
Beyond these, all the TWs are parametrised by their axial wavenumber
$\alpha$ which is a continuous variable with a finite range at a
given $Re$. These waves are exact solutions of the Navier-Stokes
equations when periodic flow conditions are imposed across the pipe
with a spatial period equal to an integer multiple of  $2
\pi/\alpha$. The 3 TWs found recently by Duguet et al (2008a) are
then introduced in section 3. Numerical codes
developed in Wedin \& Kerswell (2004) are used to explore these
waves in parameter space. A new classification is then introduced in
section 4 stimulated by the discoveries of related families. A final
discussion follows in section 5.

\section{Travelling wave symmetries}

The original work by Faisst \& Eckhardt (2003) and Wedin \& Kerswell
(2004) used the homotopy approach developed by Waleffe (1997) to
look for exact steady solutions in an appropriately chosen Galilean
frame. In this, an artificial body forcing is added to the
Navier-Stokes equations in order to drive streamwise rolls. These
advect the underlying shear towards and away from the pipe wall,
thereby generating positive and negative streamwise anomalies in the
mean flow called `streaks'. If these streaks are of sufficient
amplitude they become inflexionally unstable to streamwise-dependent
flows which if allowed to grow to sufficient amplitude can replace
the artificial forcing as the energy source for the rolls. By
choosing the forcing functions to have  certain discrete rotational
symmetries defined by
\begin{equation}
\R_m: \, (u,v,w,p)(s,\phi,z)  \rightarrow
(u,v,w,p)(s,\phi+2\pi/m,z), \label{R}
\end{equation}
and selecting streamwise-dependent instabilities having the
shift-\&-reflect symmetry
\begin{equation}
\S: \, (u,v,w,p)(s,\phi,z)  \rightarrow  (u,-v,w,p)(s,-\phi,z+\pi/\alpha),
\label{S}
\end{equation}
where $2\pi/\alpha$ is the wavelength, exact TWs were found by
continuing back to zero forcing in $m=2,3,4,5$ and $6$ discrete
rotational symmetry subspaces. All appear through saddle node
bifurcations with the lowest at $Re=1251$ corresponding to a $\R_3$
symmetric TW. Each TW family is parameterised by its axial
wavenumber $\alpha$ which, at a given $Re$, has a finite range (e.g
see figures 1 \& 2 of Kerswell \& Tutty 2007).

Later, motivated by a chaotic flow state found on the
laminar-turbulent boundary by Schneider et al (2007b), Pringle and
Kerswell (2007) generalised the forcing function to have no
rotational symmetry (see also Mellibovsky \& Meseguer 2007). As a
result, they found an asymmetric TW (formally the missing $\R_1$
state possessing one pair of streamwise rolls) with $\S$ symmetry.
This TW originates through a symmetry-breaking bifurcation from a
state which satisfies the additional shift-\&-rotate symmetry
\begin{equation}
{\bf \Omega_m}: \, (u,v,w,p)(s,\phi,z)  \rightarrow
(u,v,w,p)(s,\phi+\pi/m,z+\pi/\alpha)
\end{equation}
with $m=1$ and is therefore mirror-symmetric about $\phi=\pi/2$ (the
plane $\phi=0$ being set by the shift-\&-reflect symmetry). For
general $m$,
\begin{equation}
\S {\bf \Omega_m}^j=\Z_{\frac{j \pi}{2m}} \qquad j=1,3,5,\ldots 2m-1
\end{equation}
where ${\bf \Omega_m}^j$ implies ${\bf \Omega_m}$ is applied $j$
times and
\begin{equation} \Z_\psi: \, (u,v,w,p)(s,\phi,z) \rightarrow
(u,-v,w,p)(s,2\psi-\phi,z)
\end{equation}
represents reflection in the plane $\phi=\psi$. Helical
generalisations, which satisfy a modified shift-\&-rotate symmetry
\begin{equation}
{\bf \Omega_1^\beta}: \, (u,v,w,p)(s,\phi,z)  \rightarrow
(u,v,w,p)(s,\phi+(1+\frac{\beta}{\alpha})\pi,z+\pi/\alpha)
\end{equation}
(where $\beta$ is the helicity, Pringle \& Kerswell 2007) but no
$\S$-symmetry, were also found but always at higher $Re$ indicating
that the flow prefers the streaks to be aligned with the flow rather
than twisted around it.

\begin{figure}
 \begin{center}
 \setlength{\unitlength}{1cm}
  \begin{picture}(11.5,11.5)
  \put(0.0,0.0){\epsfig{figure=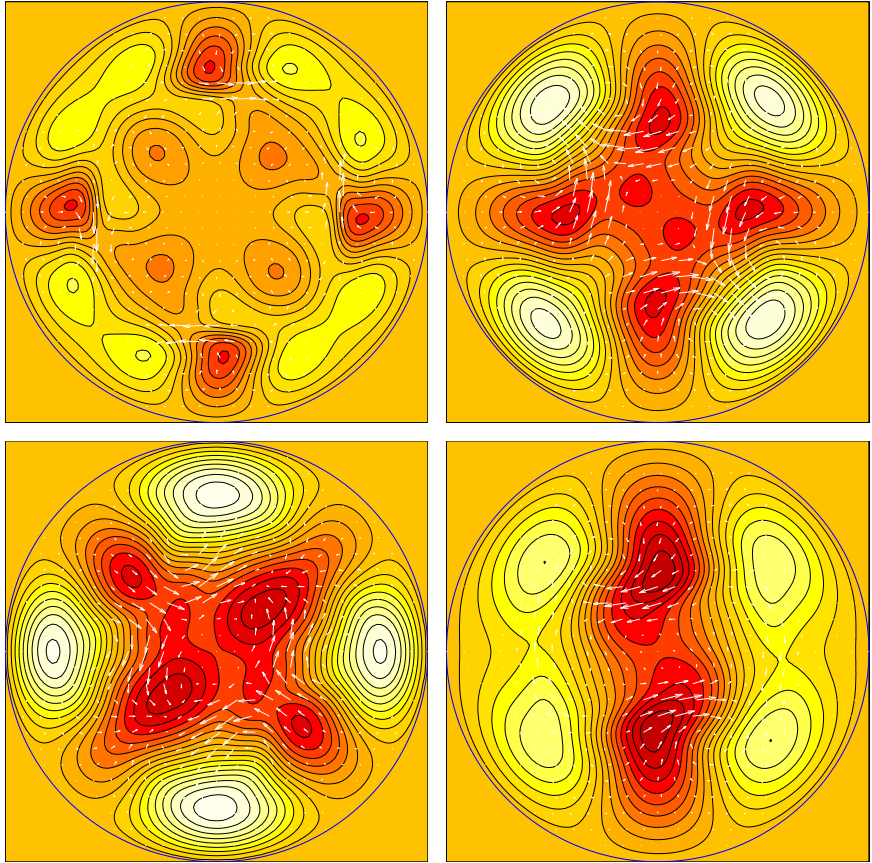,width=11.5cm,height=11.5cm,clip=true}}
  \end{picture}
  \caption{\label{fig:slices}
  The travelling waves  $A3$ (top left), $C3$ (top right)
  and $Z2$ (bottom left). For comparison purposes we
  also include $S2$ ($2b\_1.25$ in Kerswell \& Tutty 2007).
  All are shown at their respective values of $\alpha^*$ (see Table 1) and at $Re=2400$
  except for Z2 which is shown at $Re=3860$. The contours indicate the magnitude of the
  streamwise velocity difference from the underlying laminar flow. The contours and colouring
  is standardised across the plots (and more generally figures \ref{M}, \ref{N} and \ref{upper})
   with contours increments of
  $0.02U$ running from $-0.19U$(dark red) to
$0.19U$(white) (the
  colour outside of the pipes indicates zero). Arrows indicate
  the cross-stream velocities.}
  \end{center}
\end{figure}

\section{Travelling waves found on the laminar-turbulent boundary}

The new TWs discovered by Duguet et al (2008a) were found by
studying the flow dynamics on the laminar-turbulent boundary {\em
within} the $\R_2$-symmetric subspace. Surprisingly, at $Re=2875$
and periodic pipe length $ 5\,D$, the rotationally-unrestricted
situation only ever revealed the already-known asymmetric TW
(Pringle \& Kerswell 2007)  and a weakly rotating version of it
(A$^{'}$ in  Duguet et al 2008a). The fact that new $\R_2$-symmetric
TWs were observed as transient coherent structures only in the
$\R_2$-symmetric subspace calculations is due to the reduced number
of unstable directions they have as non $\R_2$-symmetric flows are
removed from the dynamics. The chance of the flow `visiting' a TW in
phase space is presumably related to how unstable it is (the spatial
periodicity of the flow was also shortened to improve the stability
of the TWs). In the notation of Duguet et al (2008a) the 3 new TWs
were labelled as `A3', `C3' and `D2' and  all are mirror symmetric
even though this symmetry was not imposed on the flow. A3 and C3
were readily converged in the continuation codes of Wedin \&
Kerswell (2004) but D2 could not be, due, we suspect, to
insufficient axial resolution being achievable in the continuation
code. The roll structure of D2 was, however, used to design a
forcing in the homotopy approach which successfully yielded a
similar-looking TW - called Z2 - with the same symmetries as D2.
That they were, in fact, different waves became apparent when Z2
could not be continued below $Re=3250$ whereas D2 was discovered at
$Re=2875$. The structure of A3, C3 and Z2 is shown in figure
\ref{fig:slices} compared to a known TW in S2, which, at
$\alpha=1.25$ and $Re=2400$ is already known to be on the
laminar-turbulent boundary (this is $2b\_1.25$ in the nomenclature
of Kerswell \& Tutty 2007).

While all the new families possess the apparently universal features
of exact coherent structures known in wall-bounded shear flows -
wavy streaks with staggered quasistreamwise vortices (Nagata 1990,
Waleffe 1998, 2001, 2003, Faisst \& Eckhardt 2003, Wedin \& Kerswell
2004, Pringle \& Kerswell 2007), there are new structural features.
A3 has a strikingly different cross-sectional profile from
previously known TWs (e.g. S2) in that {\em both} fast and slow
streaks are concentrated at the pipe wall leaving the interior
relatively quiescent. Figure \ref{fig:slices} shows that C3 and Z2
are also noticably different too. However, it is the axial structure
of Z2 which really makes it stand out. Z2 is not $\S$-symmetric nor
${\bf \Omega}$-symmetric but {\em does} have mirror symmetry about two
perpendicular planes. There is no \emph{a priori} reason to expect
the flow to prefer one of these symmetries over any of the others,
but to date all TWs in \emph{any} of the canonical shear flows have
always been \S-symmetric (except, trivially, the helical modes of
PK07; Waleffe, private communication). Therefore Z2 and D2 are the
first TWs found to possess only \Z-symmetry ($\Z_\psi$ with $\psi$
suppressed as there is no longer an origin for $\phi$). Both Z2 and
D2 are, however, close to being $\S$-symmetric in the sense that the
simple indicator
\begin{equation}
\frac{\int_0^{2\pi}\int_0^1 (\,\textbf{u}-\S \textbf{u}\,)^2|_{z=0}
s \,\textrm{d}s\,\textrm{d}\phi} {\int_0^{2\pi}\int_0^1
(\,\textbf{u}+\S \textbf{u}\,)^2|_{z=0} s
\,\textrm{d}s\,\textrm{d}\phi} =O(10^{-3}).
\end{equation}

In contrast, A3 and C3 are both $\S$- and ${\bf \Omega_2}$-symmetric
and hence also reflectionally symmetric about a plane  at $\pm \pi
/4$ to the plane of shift-\&-reflect symmetry (horizontal in figure
\ref{fig:slices}) i.e. they have  $\Z_{\pm \pi/4}$ symmetry. The
modes C3 and A3 appear through saddle node bifurcations at $Re=1141$
and $1125$ with optimal wavenumbers $\alpha^*=1.2$ and $2.0$
respectively giving these  lowest saddle node bifurcations: see Table 1.

\begin{figure}
 \begin{center}
 \setlength{\unitlength}{1cm}
  \begin{picture}(13.02,11.1)
  \put(0,0){\epsfig{figure=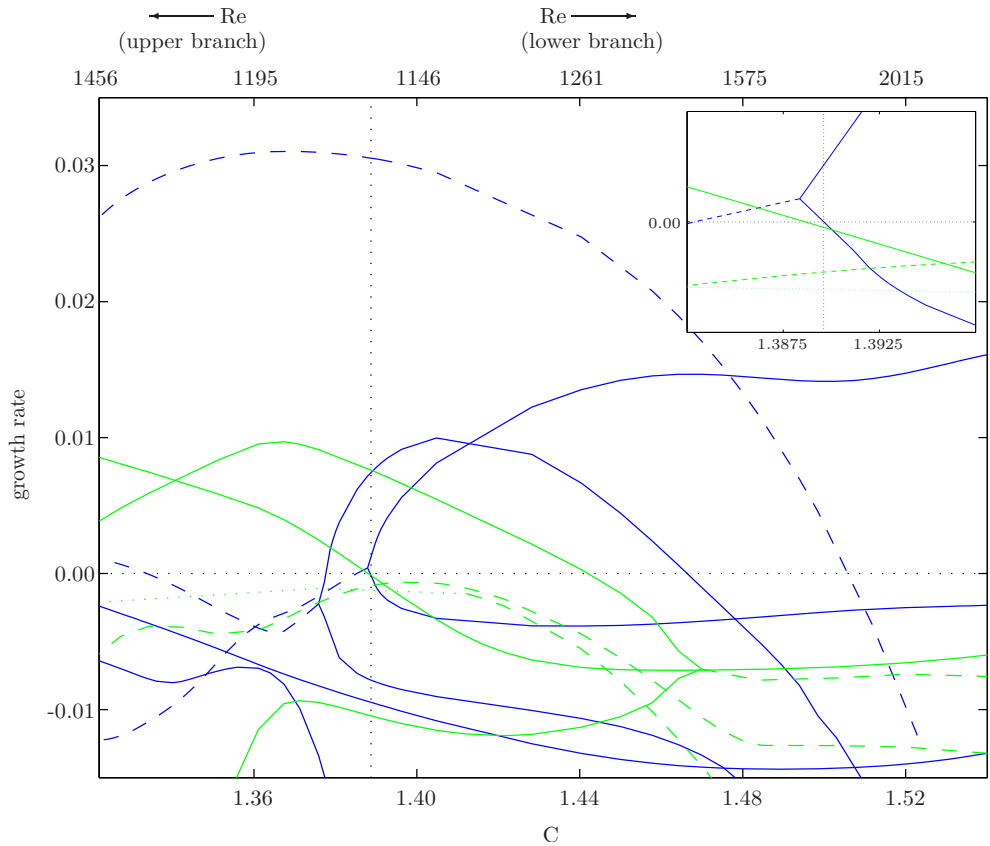,width=13.02cm,height=11.1cm,clip=true}}
  \end{picture}
  \end{center}
  \caption{\label{fig:stab} Stability of $C3$ within the $\R_2$-symmetric
  subspace against the phase speed $C$ as $Re$ increases away from the saddle node
  bifurcation at $1141$ ($\alpha=\alpha^*$ and the corresponding $Re$ is
  across the upper x-axis). Each line either indicates the locus of a real
  eigenvalue (solid) or a complex conjugate pair (dashed) as $Re$
  changes. The blue (dark) lines correspond to those which are symmetric under $\S$
  while the green (light) lines are anti-symmetric under $\S$. The vertical
  dotted line indicates the saddle node, of which the inset is a close up.
  (The dotted green indicates a stable complex conjugate pair which was difficult to resolve). }
\end{figure}

\subsection{Stability}

An interesting feature of all TWs found so far is that they are
unstable but only with an unstable manifold of very small dimension
(invariably less than 10 for those checked so far - Faisst \&
Eckhardt 2003, Kerswell \& Tutty 2007). The typical situation is
illustrated in figure \ref{fig:stab} which displays the spectrum of
the C3 wave at the wavenumber $\alpha^*$ which gives its lowest
saddle node bifurcation. While $Re$ is double-valued near the
bifurcation, the phase speed $C$ is not, monotonically decreasing in
value as the bifurcation point is crossed  from lower to upper
branch. This then provides a convenient abscissa to show how the
stability changes on both upper and lower branches as $Re$ increases
away from the saddle node bifurcation. Only disturbances which share
the $\R_2$-symmetry of C3 are considered. C3 is particularly
interesting for two reasons: 1) the number of unstable directions
decreases down to one for $Re \geq 1826$ along the lower branch and
2) there are several bifurcations involving real eigenvalues. The
implication of the first observation is that C3, which is on the
laminar-turbulent boundary (in a pipe 2.5D long at $Re=2400$, Duguet
et al 2008a), will become a local attractor there beyond $Re=1826$
for $\R_2$-symmetric flow (confirmed by Duguet et al. 2008a). The
second observation means that new solutions bifurcate from C3 which
are also steady in an appropriately translating Galilean frame -
that is to say they are travelling waves (this is how the
asymmetric travelling wave of Pringle \& Kerswell $2007$ appears).
There are 4 possibilities: a transcritical bifurcation where no
symmetry is broken, and 3 types of symmetry-breaking pitchfork in
which only $\S$-, $\Z$- or ${\bf \Omega_2}$-symmetry is retained
(recall 2 symmetries imply the third). For example, the bifurcation
at $(C,Re)=(1.47,1449)$ is a mirror-symmetry-breaking pitchfork
which can be followed to produce TWs which resemble those of S2. 
D2 is plausibly the product of a
$\S$-symmetry-breaking pitchfork from C3 (D2 is very similar to Z2
and hence C3 - see figure \ref{fig:slices}).

Hopf bifurcations are the generic scenario, however, leading to more
complicated, relative periodic orbits such as that traced by Duguet
\emph{et al.} ($2008b$). Within the $Re$ range of figure
\ref{fig:stab}, C3 experiences 3 Hopf bifurcations which
all lead to relative periodic orbits with the $\Z$-symmetry broken.

%
% table of various symmetries here
%
\begin{table}
\begin{center}
\begin{tabular}{@{}cllrr@{}}
   & & & & \\ \hline
  & & &  &\\
Source      & \qquad &  Symmetries \qquad &   $Re_{lowest}$ & $\alpha^*$ \\
& & &   & \\ \hline
              &        &                          &             &       \\
FE03 \& WK04  & S2     & $\S$ \& $\R_2$           &        1358 & 1.55  \\
              & S3     & $\S$ \& $\R_3$           &        1251 & 2.44  \\
              & S4     & $\S$ \& $\R_4$           &        1647 & 3.23  \\
              & S5     & $\S$ \& $\R_5$           &        2485 & 4.11  \\
WK04          & S6     & $\S$ \& $\R_6$           &        2869 & 4.73  \\
              &  ?     & $\S$ \& ${\bf \Omega_1}$  &        3046 & 2.17  \\
              &        &                          &             &       \\
PK07          & Asymm (S1) & $\S$                     &        $\approx 820$ & $\approx 1.8$  \\
              & Mirror-symm (M1) & $\S$ \& ${\bf \Omega_1}$ &  773 & 1.44\\
              & Helical & ${\bf \Omega_1^\beta}$     &     773 & 1.44       \\
              &         &                            &     &        \\
here          & M2 (A3) & $\S$, $\R_2$ \& ${\bf \Omega_2}$ & 1125 & $2.0$\\
              & M3      & $\S$, $\R_3$ \& ${\bf \Omega_3}$ & 1552 & 2.2 \\
              & M4      & $\S$, $\R_4$ \& ${\bf \Omega_4}$ & 1824 & 2.6\\
              & M5      & $\S$, $\R_5$ \& ${\bf \Omega_5}$ & 2143 & 3.1\\
              & M6      & $\S$, $\R_6$ \& ${\bf \Omega_6}$ & 2531 & 3.5\\
              &         &                                 &      &\\
              & N2 (C3)   & $\S$, $\R_2$ \& ${\bf \Omega_2}$ & 1141 & 1.2\\
              & N3      & $\S$, $\R_3$ \& ${\bf \Omega_3}$ & 1037 & 2.0\\
              & N4      & $\S$, $\R_4$ \& ${\bf \Omega_4}$ & 1290 & 2.5\\
              & N5      & $\S$, $\R_5$ \& ${\bf \Omega_5}$ & 1622 & 2.9\\
              &         &                                  &     &\\
              & Z2      & $\Z$ \& $\R_2$                  & $\approx$ 3250& 0.8\\
              & D2      & $\Z$ \& $\R_2$                  & $<\,$2875 &? \\
              &         &                    &            &    \\ \hline
              &         &                    &            &
\end{tabular}
\end{center}
\caption{{\rm The various symetries of all the TWs currently known.
FE04 is Faisst \& Eckhardt (2003), WK04 is Wedin \& Kerswell (2004)
and PK07 is Pringle \& Kerswell (2007). The '?' mark in the entry
for WK04 is to indicate that they found a mirror-symmetric  TW but
at high $Re$ and its relationship to M1 is unclear. }}
\label{table1}
\end{table}

\begin{figure}
 \begin{center}
 \setlength{\unitlength}{1cm}
  \begin{picture}(11,11) \put(0,0){\epsfig{figure=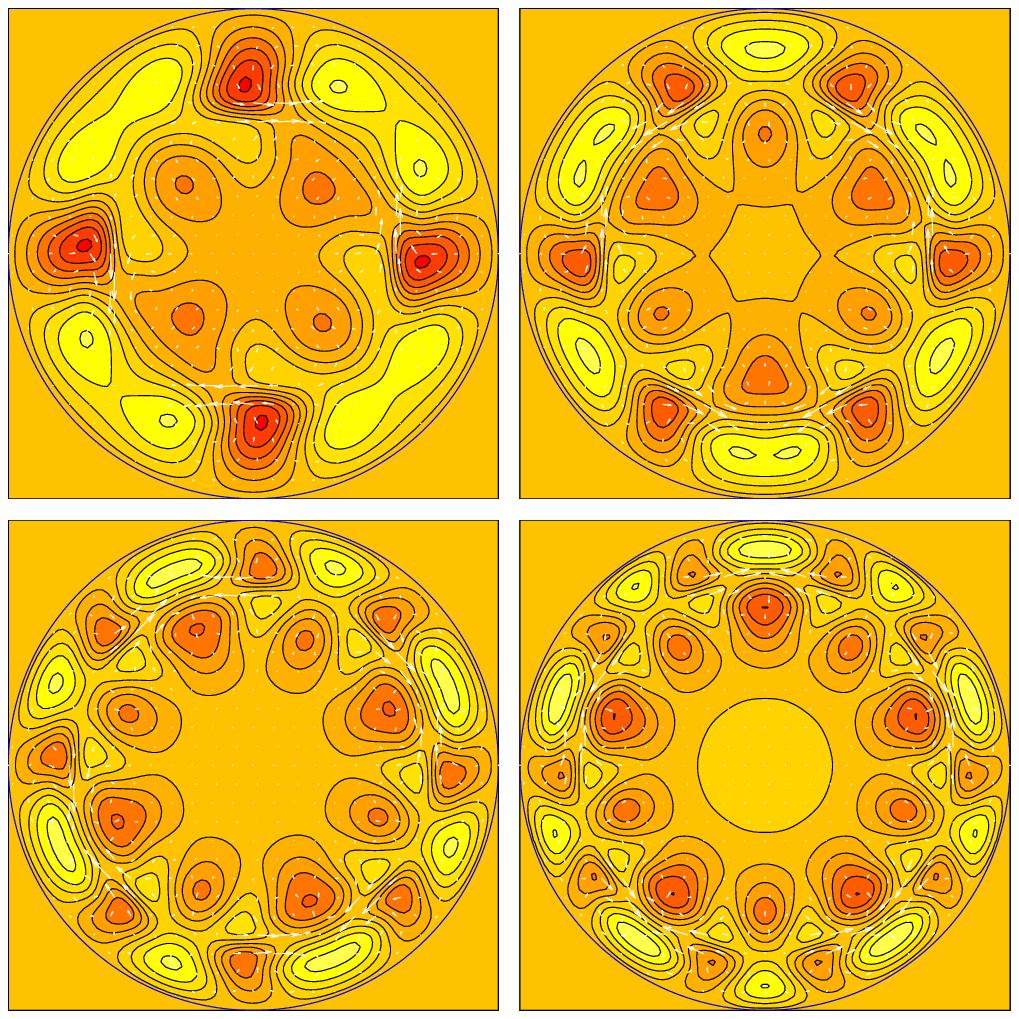,width=11cm,height=11cm,clip=true}}
  \end{picture}
  \end{center}
  \caption{\label{M} Slices of the lower branch M-class TW solutions at $Re=2400$.
  M2 (A3,top left), M3 (top right), M4 (bottom left) and M5 (bottom right).
  Contour levels of the streamwise velocity perturbation are in
  increments of $0.02U$ running from $-0.19U$(dark red) to $0.19U$(white). }
\end{figure}

\begin{figure}
 \begin{center}
 \setlength{\unitlength}{1cm}
  \begin{picture}(11,11) \put(0,0){\epsfig{figure=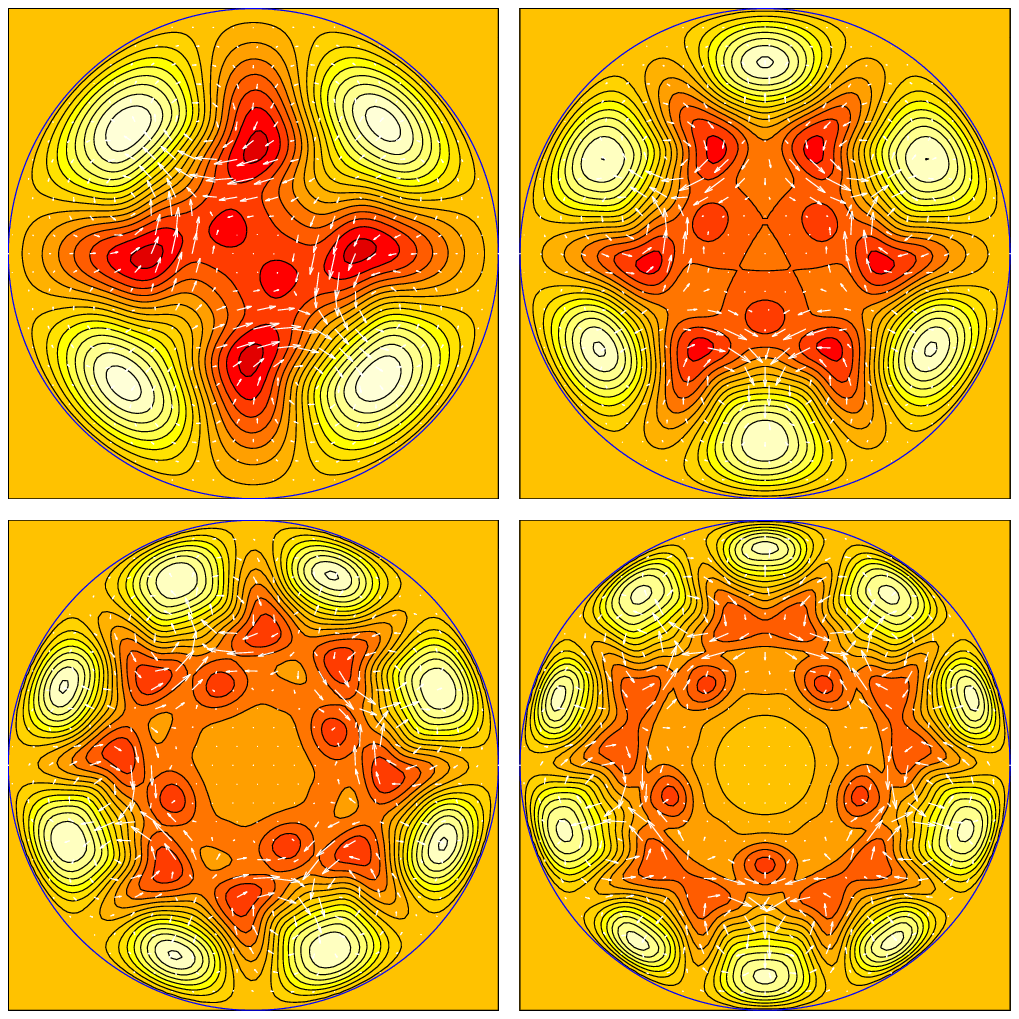,width=11cm,height=11cm,clip=true}}
  \end{picture}
  \end{center}
  \caption{\label{N} Slices of the lower branch N-class TW solutions at $Re=2400$.
  N2 (C3, top left), N3 (top right), N4 (bottom left) and N5 (bottom right).
  Contour levels of the streamwise velocity perturbation are in
  increments of $0.02U$ running from $-0.19U$(dark red) to $0.19U$(white).}
\end{figure}

\begin{figure}
 \begin{center}
 \setlength{\unitlength}{1cm}
  \begin{picture}(13.44,10.11)
  \put(0,0){\epsfig{figure=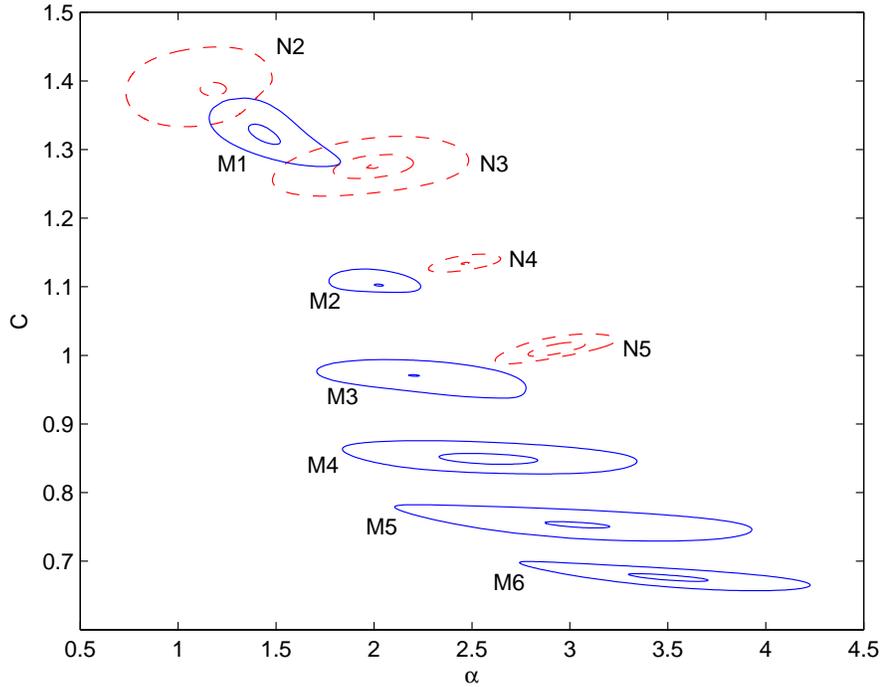,width=13.44cm,height=10.11cm,clip=true}}
  \end{picture}
  \end{center}
  \caption{\label{Calpha}
  A plot of the phase speed $C$ (in units of $U$) against the axial wavenumber $\alpha$
  (in units of $2/D$).
  The contours correspond to different $Re$ with the loops
  (M-class/N-class TWs shown using blue solid/red dashed lines)
  constricting as they move towards  the saddle nose
  indicating $\alpha^*$. The various $Re$
   for each TW are as follows: M1 776 \& 820, M2 1125 \& 1145, M3 1552 \& 1650,
    M4 1846 \& 2037, M5 2148 \& 2400,
  M6 2540 \& 2662; N2 1150 \& 1318, N3 1038, 1050 \& 1120, N4 1292 \& 1300, N5 1629 \& 1652.}
\end{figure}

\begin{figure}
 \begin{center}
 \setlength{\unitlength}{1cm}
 \begin{picture}(13.44,10.11)
 \put(0,0){\epsfig{figure=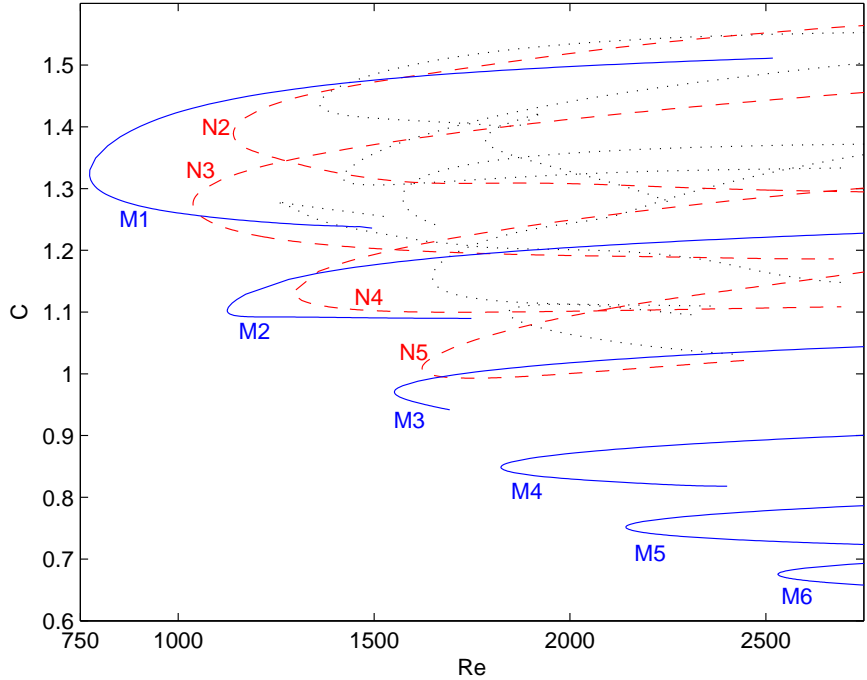,width=13.44cm,height=10.11cm,clip=true}}
  \end{picture}
  \end{center}
  \caption{\label{CRe}
  A plot of $C$ against $Re$ for $\alpha^*$ (see Table 1) which gives the lowest saddle node
  bifurcation for each TW (M-class/N-class waves shown using blue
  solid/red dashed lines). The non-mirror
  symmetric S-class of Faisst \& Eckhardt(2003) and Wedin \& Kerswell (2004) which
  generally appear at higher $Re$ are shown using black dotted lines. Typical resolutions used
  (in the nomenclature of Wedin \& Kerswell (2004)) are: M1 (14,25,5), M2 (12,30,6), M3 (10,30,10),
  M4 (10,30,8), M5 \& M6 (10,35,8); N2 (12,30,6), N3 \& N4 (10,30,8), N5 (8,35,8).}
\end{figure}

\begin{figure}
 \begin{center}
 \setlength{\unitlength}{1cm}
  \begin{picture}(11,11) \put(0,0){\epsfig{figure=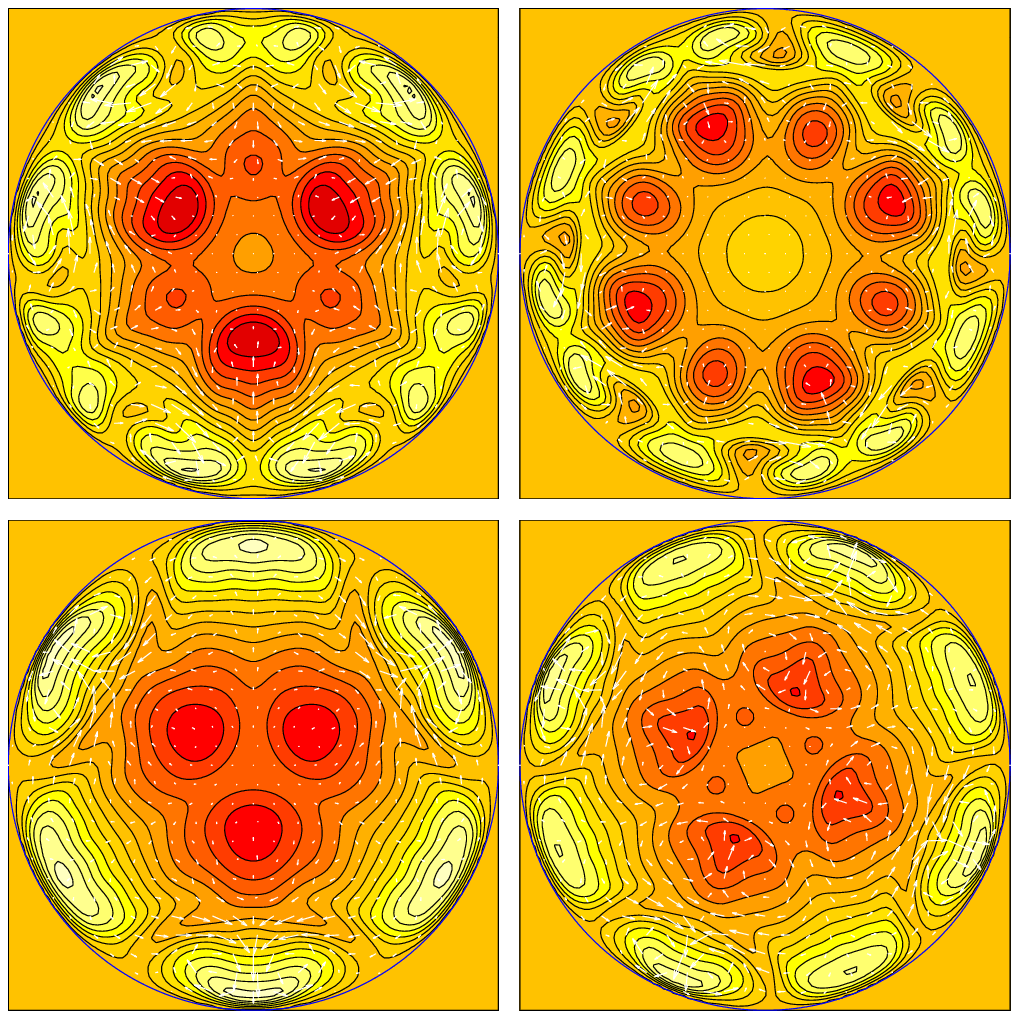,width=11cm,height=11cm,clip=true}}
  \end{picture}
  \end{center}
  \caption{\label{upper} Slices of upper branch solutions at $Re=2400$.
  M3 (top left), M4 (top right),
  N3 (bottom left) and N4 (bottom right). Contour levels as in figures \ref{M} and \ref{N}.}
\end{figure}

\section{New mirror-symmetric classes}

The fact that A3 and C3 appear at such low $Re$ strongly suggests
that there are analogous waves in different rotational symmetry
classes also existent at pre- and transitional $Re \lesssim 2400$.
This indeed turns out to be the case with all except one (M3) of the
new families of mirror-symmetric modes being easily found using the
homotopy approach once the appropriate ${\bf \Omega}$-symmetry is
incorporated. Selecting a roll forcing of $\R_{2m}$-symmetry
defined, in the notation of Wedin \& Kerswell ($2004$), by
$\lambda_{2m\,i}$ with $i=2,3$ or $4$ invariably led to a
subharmonic $\R_m$-symmetric streak instability which could be
easily tracked back to zero forcing (this was exactly the strategy
used by Wedin \& Kerswell to find their $\R_1$ TW which has two roll
pairs and appears beyond $Re=3046$). The remaining wave M3 was found
by homotopy in $m$ starting at a M5 wave (M5 being used rather than
M4 because of the similar parity of velocity components with respect
to the radius).

Inspection of the new lower branch TWs - see Figures \ref{M} and
\ref{N}\footnote{Videos to be included
  with this paper are particularly illuminating -  they are currently on  http://www.maths.bris.ac.uk/$\sim$cp1571/TWs/table.htm} - indicates that the
mirror-symmetric family reported in Pringle \& Kerswell (2007) is
the first family (M1) and A3 is a member of the second family (M2)
of a class of TWs (M$m$ with $m$=2,3,4,5,6, \ldots) with two layers
of fast and slow streaks.  Furthermore, C3 appears a member of a
$\R_2$-symmetric family (N2) of another class of TWs (N$m$ with
$m$=2,3,4,5,\ldots). The ordering of the families within the
respective classes is clear from the phase speed and wavenumber data
shown in figure \ref{Calpha}.

Extending the M$m$ and N$m$ families to higher rotational symmetry
$m$ is straightforward and could have been continued if the general
trend had not already emerged. Finding lower families is, however,
more difficult with N1 noticeably absent at the time of writing.
This family could potentially be the first to appear as $Re$
increases from zero but this seems less likely after plotting the
optimal slices of each family's solution surface together on a phase
speed verses $Re$ plot (`optimal' being defined by that wavenumber
$\alpha^*$ which corresponds to the lowest saddle node bifurcation).
Whereas the bifurcation points for M$m$ TWs monotonically increase
in $Re$ and decrease in $C$ with $m$ (see figure \ref{CRe}), 
this is not true for the
N-class. Of N2 to N5, N3 actually has the lowest bifurcation point.
This mimics the situation for the original S-class (non-mirror
symmetric) TWs. The fact that the lowest $Re$ for each N$m$ family
as well as M1 and M2 is consistently smaller than the lowest $Re$
for the equivalent S$m$ family is suggestive that the latter all
bifurcate off the former in mirror-symmetry-breaking bifurcations.
This, after all, is the situation for the asymmetric waves reported
in Pringle \& Kerswell (2007) which form the missing (trivially)
$\R_1$-symmetric family from the original studies and which
bifurcate off the M1 family. For a given $(\alpha,Re)$, a TW needs
at least 2 quantities such as the the phase speed $C$ and a Reynolds
number, $Re_p$, measuring the wall shear stress to characterise it.
Examining N3 and S3 TWs shows that they occur in similar parts of
$(Re,C,Re_p)$ space at $\alpha=2.44$ ($\alpha^*$ for S3) but there
is no obvious connection between them at least for $Re \leq 2500$.
The implication then is, if S3 does bifurcation off N3, the
solutions exist to  significantly subcritical $Re$ which would
explain why the importance of mirror-symmetric TWs hadn't been
realised before.

%
% friction factor plot
%

\begin{figure}
 \begin{center}
 \setlength{\unitlength}{1cm}
  \begin{picture}(11.85,8.88)
  \put(0,0){\epsfig{figure=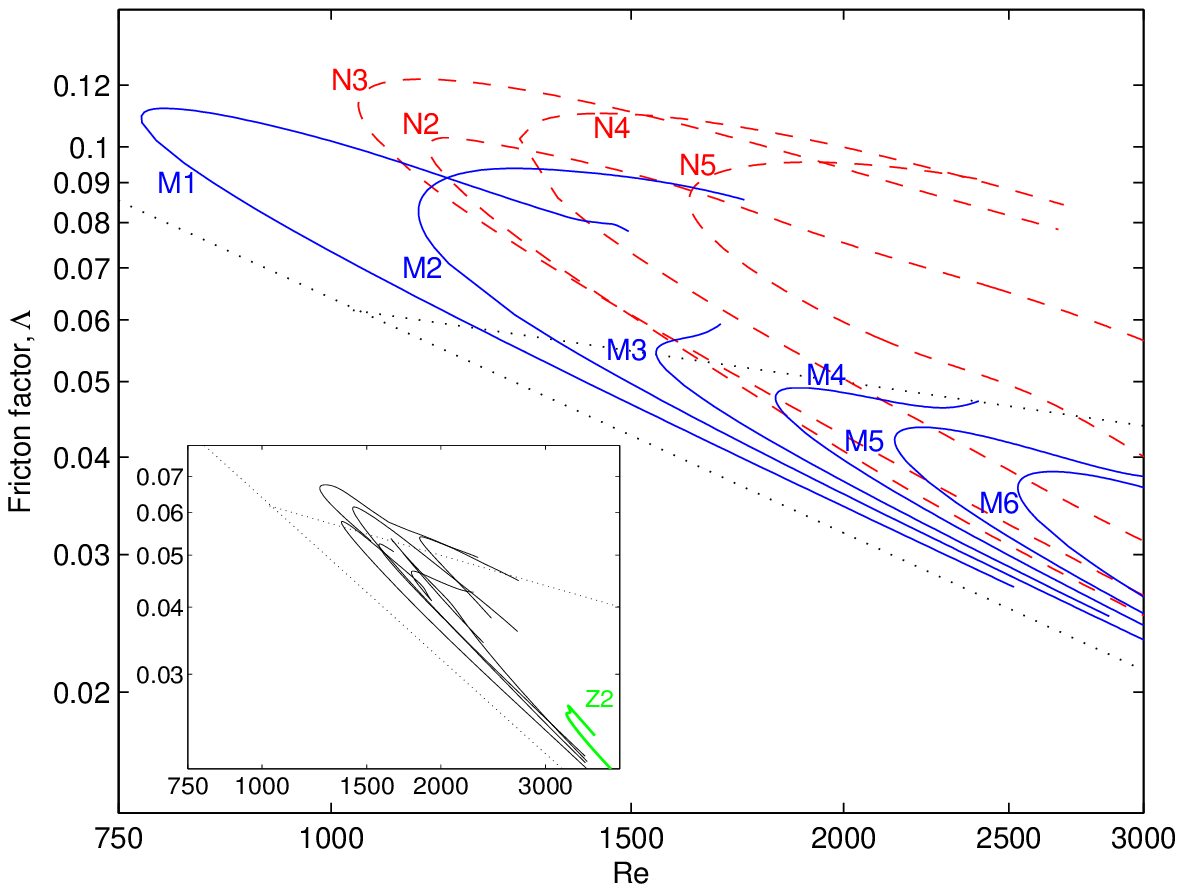,width=11.85cm,height=8.88cm,clip=true}}
  \end{picture}
  \end{center}
  \caption{\label{fig:fric}
    The friction factor, $\Lambda$, against Re for the newly found
    travelling waves, as well as the asymmetric wave (M1). The lower dotted
    line corresponds to the laminar state, and the upper dotted line
    to the log-law parameterisation of experimental data, $1/\sqrt
    {\Lambda}=2.0\log(Re_m\sqrt{\Lambda})-0.8$ (see Schlichting 1968, equation (20.30)).
    The inset shows the same
    plot but for the travelling waves S2$-$S6 (only selected curves drawn for clarify) and Z2 which
    appears at much higher $Re$. The
    earlier onset in $Re$, and significantly higher friction factors
    of the new $M$ and $N$-class travelling waves is clearly
    apparent.}
\end{figure}

The new upper branch TW solutions - see figure \ref{upper} - all
show an expected intensification of the slow and fast streaks, and
the positioning of the fast streaks closer to the pipe wall than the
corresponding lower branch solutions. This localisation is
particularly noticeable for the N-class TWs indicating high wall
shear stresses which is borne out by figure \ref{fig:fric}. The
friction factors achieved by the upper branch and even by some of
the lower branch N-class TWs as well as M1 and M2 are significantly
higher than time-averaged experimental values (as shown by the upper
dotted line) and those of the S-class TWs.

Finally, it's worth remarking that the M-class TWs start to appear
at fascinatingly low $Re$ given their intricate double radial layer
structure: for example, M4 appears at $Re =1824$ (see figure
\ref{CRe} and Table 1). That they exist at all is not a surprise
(presumably TWs with three radial layers are possible too), but that
they appear so early in $Re$ surely is and contrasts starkly with Z2
which doesn't emerge until $Re \approx 3250$.

\section{Discussion}

In this paper we have described 2 new classes of mirror-symmetric
TWs - the M- and N-classes which are also both shift-\&-reflect
symmetric - and a new mirror-symmetric family Z2  which is not. The
M- and N-class waves appear earlier in $Re$ than the original
non-mirror-symmetric waves of Faisst \& Eckhart (2003) and Wedin \&
Kerswell (2004) suggesting that the latter are borne through
generally supercritical symmetry-breaking bifurcations from them.
This was already found to be the case for the M1 waves (Pringle \&
Kerswell 2007) but now seems more generally true now that the
various TW families making up each class have been found. The
stability analysis of N2 presented here provides a timely reminder
of the bifurcation possibilities: $\S$ or $\Z$ symmetries can be
broken at pitchfork bifurcations leading to either Z-class TWs or
$S$-class TWs found originally (Faisst \& Eckhardt 2003, Wedin \&
Kerswell 2004). Hopf bifurcations, of course, give rise to periodic
orbits in the Galilean frame  of the TW or quasiperiodic orbits in
the pipe (laboratory) frame. Tracing these requires a more
sophisticated approach based upon time-stepping the equations and
searching for an exact return of an appropriately chosen
Poincar\'{e} map (Viswanath 2007,2008, Duguet et al 2008b).

The original motivation for searching for all the new TWs discussed
here was the discovery of previously-unknown TWs in the
laminar-turbulent boundary by Duguet et al (2008a). It is likely
that some of the lower branch solutions of these new waves similarly
populate this boundary so that their stable manifolds also play a
role in determining if a given initial condition will lead to
relaminarisation or a turbulent evolution. There are, however, many
interesting issues surrounding this assumption. For example, if a
lower branch TW is on the boundary at one $Re$, will it still be at
$10Re$ and if not, how did it leave? In a long pipe where the TWs
are continuously parametrised by their wavenumber over a finite
range, at what critical wavenumber does the lower branch TW leave
the boundary on its way to becoming an upper branch TW? The
asymmetric wave (Pringle \& Kerswell 2007) provides an obvious
example being on the boundary for $\alpha=0.625$ (at $Re=2875$,
Schneider et al 2007b, Duguet et al 2008a) but presumably not for
$\alpha=1.44$ where its wall shear stress is high (see figure 6 of
Pringle \& Kerswell 2007). Hopefully these issues will be discussed
elsewhere in this celebratory volume.

The discovery that the upper branch solutions of the new N-class
waves have such high wall shear stresses is, however, potentially
more important. A recent attempt to extract coherent fast-streak
structures from pipe turbulence (Willis \& Kerswell 2008) has
concluded that the TWs currently known were not energetic enough to
be part of the turbulent attractor as previously supposed and
speculated about others as-yet-undiscovered. The new N-class of
mirror-symmetric TWs may well be these missing waves and the issue
clearly needs to be revisited.

In this report, we hope a step forward has been made in appreciating
the hierarchy of TWs which exist in pipe flow. Generally the picture
is that TWs with shift-\&-reflect symmetry, mirror-symmetry and a
low degree of rotational symmetry seem to appear first and spawn
further, less symmetric, TWs through bifurcations as $Re$ increases.
However, given the notorious complexity of the Navier-Stokes
equations, it would be foolhardy to be too rigid about this
especially as the N1 family still remains `at large'. What should be
absolutely clear, though, is that the pipe flow problem continues to
fascinate and intrigue us fully 125 years after Reynolds' original
experiments.

\begin{acknowledgements}
CP acknowledges studentship support from EPSRC and YD the award of a
Marie-Curie IntraEuropean Fellowship (grant number
MEIF-CT-2006-024627).
\end{acknowledgements}

\end{document}